\documentclass[12pt,fleqn]{article}
\usepackage{amsmath,amssymb}
\usepackage{bm}
\usepackage[dvips]{graphicx}
\newcommand{\rth}{\frac{1}{\sqrt{3}}}
\newcommand{\rsix}{\frac{1}{\sqrt{6}}}

\allowdisplaybreaks[3]

\begin{document}

\date{}

\title{  On the determination of $\Theta^+$ quantum numbers and other topics of
exotic baryons}

\author{E.~Oset$^a$,T.~Hyodo$^b$, A.~Hosaka$^b$, F. J. Llanes-Estrada$^c$, V. Mateu$^a$,
S. Sarkar$^a$ and\\
 M.J. Vicente Vacas$^a$}

\maketitle

\begin{center}

$^a$\textit{Departamento de F\'\i sica Te\'orica and IFIC,
Centro Mixto Universidad de Valencia-CSIC,
Institutos de Investigaci\'on de Paterna, Aptd. 22085, 46071
Valencia, Spain}\\
$^b$\textit{Research Center for Nuclear Physics (RCNP),
Ibaraki, Osaka 567-0047, Japan}\\ 
$^c$\textit{Deapartamento de Fisica Teorica I, Universidad Complutense, Madrid,
Spain}\\
\end{center}

\section{Introduction}
In this talk I look into three different topics, addressing first a method to 
determine the quantum numbers of the $\Theta^+$, then exploiting the possibility
that the $\Theta^+$ is a bound state of $K \pi N$ and in the third place I
present results on a new resonant exotic baryonic state which appears as
dynamically generated by the Weinberg Tomozawa $\Delta K$ interaction. 

\section{Determining the $\Theta^+$ quantum numbers
through the $K^+p\to \pi^+K^+n$}
A recent experiment by LEPS collaboration
at SPring-8/Osaka~\cite{nakano}
has found a clear signal for an $S=+1$ positive charge resonance
around 1540 MeV.
The finding, also confirmed by DIANA at
ITEP~\cite{Barmin:2003vv},
CLAS at Jefferson Lab.~\cite{Stepanyan:2003qr}
and SAPHIR at ELSA~\cite{Barth:2003es} and other more recent experiments,
might correspond to the exotic state predicted by Diakonov {\em et~al.}
in Ref.~\cite{Diakonov:1997mm}, but since then much theoretical work has been
done to understand the nature of this resonance, see \cite{elena} for a recent
review of theoretical and experimental work done. Yet, the spin, parity and isospin are not determined
 experimentally. We present here one particularly suited reaction to determine 
 the quantum numbers with the process
\begin{equation}
    K^+p\to \pi^+K^+n \ .
    \label{eq:reaction}
\end{equation}

\begin{figure}[tbp]
    \centering
    \includegraphics[width=10cm,clip]{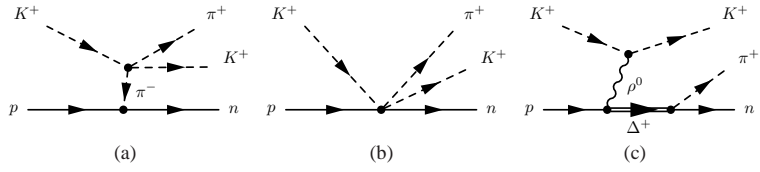}
    \caption{\label{fig:1}
    Feynman diagrams of the reaction $K^+p\to \pi^+K^+n$
    in the model of Ref.~\cite{Oset:1996ns}.}
\end{figure}

\begin{figure}[tbp]
    \centering
    \includegraphics[width=10cm,clip]{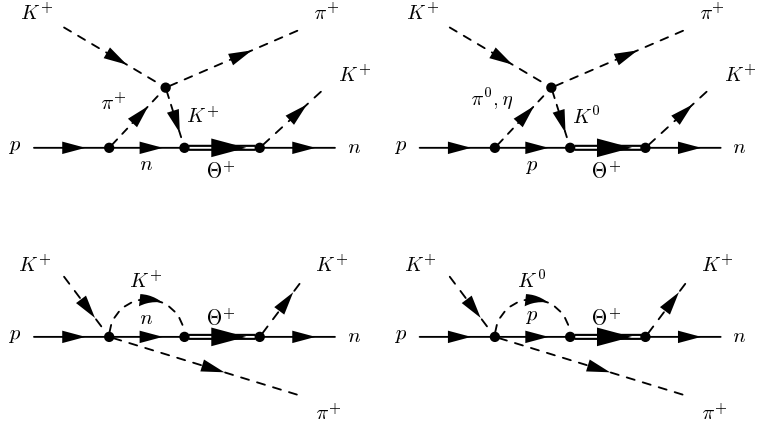}
    \caption{\label{fig:2}
    Feynman diagrams of the reaction $K^+p\to \pi^+K^+n$
    with the $\Theta^+$ resonance.}
\end{figure}

A successful model for the reaction~\eqref{eq:reaction} was
considered in
Ref.~\cite{Oset:1996ns}, consisting of the mechanisms depicted in
terms of Feynman diagrams in Fig.~\ref{fig:1}.
The term (a) (pion pole) and (b) (contact term), which are easily
obtained from the chiral Lagrangians involving
meson-meson~\cite{Gasser:1985gg}
and meson-baryon interaction~\cite{Bernard:1995dp}
are spin flip terms (proportional to $\bm{\sigma}$), while the $\rho$
exchange term (diagram (c)) contains both a spin flip and a non spin 
flip part. Having an amplitude proportional to $\bm{\sigma}$ is
important in the present context in order to have a test of the
parity of the resonance.
Hence we choose a situation, with the final pion momentum 
$\bm{p}_{\pi^+}$ small compared to 
the momentum of the initial kaon, such that the diagram (c), which 
contains the $\bm{S}\cdot \bm{p}_{\pi^+}$ operator can be safely
neglected.
The terms of Fig.~\ref{fig:1} (a) and (b) will provide the bulk for
this reaction.
If there is a resonant state for $K^+n$ then this 
will be seen in the final state interaction of this system.
This means that in addition to the diagrams (a) and (b) 
of Fig.~\ref{fig:1},
we shall have those in Fig.~\ref{fig:2}.
If the resonance is an $s$-wave $K^+n$ resonance then $J^P=1/2^-$.
If it is a $p$-wave resonance, we can have $J^P=1/2^+, 3/2^+$.
A straightforward evaluation of the meson pole and contact terms (see also
Ref.~\cite{Meissner:1995vp}) leads to the $K^+n\to \pi^+KN$
amplitudes
\begin{equation}
    -it_i
    =(a_i+b_i\bm{k}_{in}\cdot \bm{q}^{\prime}+c_i)\bm{\sigma}
    \cdot\bm{k}_{in}
    +(-a_i-b_i\bm{k}_{in}\cdot \bm{q}^{\prime}+d_i)\bm{\sigma}
    \cdot\bm{q}^{\prime} \ ,
    \label{eq:t1amp}
\end{equation}
where $i=1,2$ stands for the final state $K^+n, K^0p$ respectively
and $k_{in}$ and $q^{\prime}$ are the initial and final $K^+$ momenta.
The coefficients $a_i$ and $b_i$ are from meson exchange terms, and
$c_i$ and $d_i$ from contact terms.
They are given in Ref. \cite{hyodo}.
When taking into account $KN$ scattering through the $\Theta^+$
resonance, as depicted in Fig.~\ref{fig:2}, the $K^+p\to \pi^+K^+n$
amplitude is given by
\begin{equation}
    -i\tilde{t}=-it_1-i\tilde{t}_1-i\tilde{t}_2
    \label{eq:total}
\end{equation}
where $\tilde{t}_1$ and $\tilde{t}_2$ account for the scattering
terms with intermediate $K^+n$ and $K^0p$, respectively.
They are given by
\begin{equation}
\begin{split}
    -\tilde{t}^{(s)}_i
    =&\frac{g_{K^+n}^2}{M_I-M_R+i\Gamma/2}
    \left\{G(M_I)(a_i+c_i)
    -\frac{1}{3}\bar{G}(M_I)b_i\right\}
    \bm{\sigma}\cdot\bm{k}_{in}S_I(i) \ ,\\
    -\tilde{t}^{(p,1/2)}_i
    =&\frac{\bar{g}_{K^+n}^2}{M_I-M_R+i\Gamma/2}
    \bar{G}(M_I)
    \left\{\frac{1}{3}b_i\bm{k}_{in}^2-a_i+d_i
    \right\}
    \bm{\sigma}\cdot\bm{q}^{\prime} S_I(i) \ , \\
    -\tilde{t}^{(p,3/2)}_i
    =&\frac{\tilde{g}_{K^+n}^2}{M_I-M_R+i\Gamma/2}
    \bar{G}(M_I)
    \frac{1}{3}b_i
    \left\{
    (\bm{k}_{in}\cdot\bm{q}^{\prime})
    (\bm{\sigma}\cdot\bm{k}_{in})
    -\frac{1}{3}\bm{k}_{in}^2\bm{\sigma}\cdot\bm{q}^{\prime}
    \right\}S_I(i) \ , 
\end{split}
    \label{eq:tilde}
\end{equation}
for $s$- and $p$-wave, and $i=1,2$ for $K^+n$ and $K^0p$
respectively.
The different magnitudes of Eqs.~\eqref{eq:tilde} are defined in \cite{hyodo},
but the only thing to recall here is the dependence on the momenta of the 
$\bm{\sigma}\cdot\bm{p}$ terms.
Invariant mass distributions and angular distributions are given in \cite{hyodo}.
Here we only want to discuss the polarization obervables.

Let us now see what can one learn with resorting to polarization
measurements. 
Eqs.~\eqref{eq:tilde} account for the resonance contribution
to the process.
The interesting finding there is that if the $\Theta^+$ 
couples to $K^+n$ in $s$-wave (hence negative parity) 
the amplitude goes as $\bm{\sigma}\cdot \bm{k}_{in}$, 
while if it couples in $p$-wave it has a term
$\bm{\sigma}\cdot \bm{q}^{\prime}$.
Hence, a possible polarization test to determine which one of the
couplings the resonances chooses is to measure the cross section for
initial proton polarization $1/2$ in the direction $z$ $(\bm{k}_{in})$
and final neutron polarization $-1/2$  (the experiment
can be equally done with $K^0p$ in the final state, which makes the
nucleon detection easier).
In this spin flip amplitude $\langle -1/2 |t|+1/2\rangle$, the 
$\bm{\sigma}\cdot \bm{k}_{in}$ term vanishes.
With this test the resonance signal disappears for the $s$-wave
case, while the $\bm{\sigma}\cdot \bm{q}^{\prime}$ operator of the
$p$-wave case would have a finite matrix element 
proportional to $q^{\prime}\sin \theta$.
This means, away from the forward direction of the final kaon, the
appearance of a resonant peak in the cross section would indicate a
$p$-wave coupling and hence a positive parity resonance.

\begin{figure}[tbp]
    \centering
    \includegraphics[width=10cm,clip]{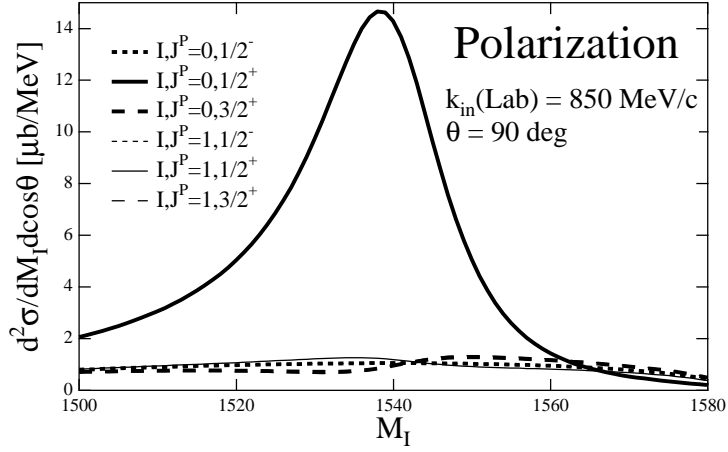}
    \caption{\label{fig:5}
    The double differential cross sections of polarized amplitude
    with $\theta=90$ for $I=0,1$ and $J^P=1/2^-,1/2^+,3/2^+$.
    }
\end{figure}

In Fig.~\ref{fig:5} we show the results for the polarized cross section
measured at 90 degrees as a function of the invariant mass.
The two cases with s-wave do not show any resonant shape
since only the background contributes.
All the other cross sections are quite reduced to the point
that the only sizeable resonant peak comes from the $I,J^P=0,1/2^+$ case.
A clear experimental signal of the resonance in this observable
would unequivocally indicate the quantum numbers as  $I,J^P=0,1/2^+$.

\section{Is the $\Theta ^+$ a $K \pi N$ bound state?}

At a time when many  low energy baryonic resonances are being
dynamically generated as meson baryon quasibound states within chiral 
unitary
approaches \cite{Kaiser:1995eg,Oset:1997it,oller,
Garcia-Recio:2002td,Jido:2003cb,Nieves:2001wt} it 
looks tempting to
investigate the
possibility of this state being a quasibound state of a meson and a baryon 
or two  mesons and a baryon.  Its nature as a $K N$ s-wave state is easily
ruled 
out since the interaction is repulsive. 
$KN$ in a p-wave, which is attractive, is too weak to 
bind. The next logical possibility is to 
consider a quasibound state of $K 
\pi N$, which in s-wave would naturally correspond to spin-parity 
$1/2^+$, the quantum numbers suggested in
\cite{Diakonov:1997mm}.
Such an idea has already been put forward in \cite{Bicudo:2003rw} where a  
study of the interaction of the three body system is conducted in the 
context of chiral quark models, which 
suggests  that it
is not easy to bind the system although one cannot rule it out completely. A
more detailed work is done in \cite{felipe}, which we summarize here.

Upon considering the possible structure of $\Theta^+$ 
we are guided by the  experimental observation
\cite{Stepanyan:2003qr} that the
state is not produced in the $K^+ p$ final state. This would
rule out the possibility of the $\Theta$ state having 
isospin I=1. Then we accept the $\Theta^+$ to be an I=0 state. 
As we couple a pion and a kaon to the nucleon to form such state,
a consequence is that the $K \pi$ substate must combine to I=1/2 and 
not I=3/2.
 This is also welcome dynamically since the s-wave $K \pi$ interaction in
I=1/2 is attractive (in I=3/2 repulsive) \cite{Oller:1998hw0}.
The attractive interaction in I=1/2 is very strong and gives
rise to the dynamical generation of the scalar $\kappa$ resonance around 
850 MeV and with a large width \cite{Oller:1998hw0}.

One might next
question that, with such a large width of the $\kappa$, the $\Theta^+$
could not be so narrow as experimentally reported.
However, this large $\kappa$ width is no problem since in our scenario 
it would arise from $K\pi$ decay, but now the $K \pi N$ decay of the 
$\Theta^+$ is forbidden as the $\Theta^+$ mass is below the $K \pi N$  
threshold.

One might hesitate to call the possible theoretical $\Theta^+$ state a
$\kappa N$ quasibound state because of the large gap of about 200 MeV to the
nominal $\kappa N$ mass. The name though is not relevant here and we can 
opt by calling it simply a $K \pi N$ state, but the fact is that the
$K\pi$ system is strongly correlated even at these lower energies, and
since 
this favours the binding of the $K \pi N$ state we  take it into
account.

In order to determine the possible $\Theta^+$ state we search for poles of
the $K\pi N \to K\pi N$ scattering matrix. To such point we construct the
series of diagrams of fig. 4.
\begin{figure}[tbp]
    \centering
    \includegraphics[width=13cm,clip]{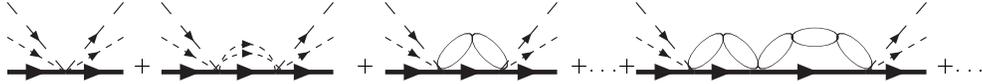}
\caption{\label{trickedamplitude1} Diagrams considered in the $\kappa N$
interaction.} 
\end{figure}
where we account explicitly for the $K\pi$ interaction by constructing
correlated $K\pi$ pairs and letting the intermediate $K\pi$ and nucleon 
propagate. 
This requires a kernel for the two meson-nucleon interaction which we
now address. 
We formulate the meson-baryon lagrangian in terms of
the SU(3) matrices, $B$, $\Gamma_\mu$, $u_{\mu}$
 and the implicit meson matrix 
$\Phi$ standard in ChPT \cite{Bernard:1995dp},
\begin{equation}  
{\it L}= {\rm Tr} \left( \overline{B} i\gamma^\mu \nabla_\mu B \right)
- M_B {\rm Tr}\left(\overline{B}B  \right) \\
+ \frac{1}{2} D {\rm Tr} \left( \overline{B} \gamma^\mu \gamma_5 \left\{
u_\mu,B\right\}   \right) +
\frac{1}{2} F {\rm Tr} \left( \overline{B} \gamma^\mu \gamma_5 \left[
u_\mu, B \right] \right)
\end{equation}
with the definitions in \cite{Bernard:1995dp}.

First there is a contact three body force simultaneously involving 
the pion, kaon and nucleon, which can be derived from the meson-
baryon Lagrangian  term containing the covariant derivative $\nabla_{\mu}$.

Next we show that a nucleon, kaon and pion see an attractive interaction
in an isospin zero state through this contact potential. 
By taking the isospin I=1/2 $\kappa$ states
and combining them with the nucleon, also isospin $1/2$, we generate 
I=0,1 states
which diagonalize the scattering matrix associated to $t_{mB}$ 
\begin{eqnarray}  
\langle \Theta^1 \arrowvert t_{mB}^s \arrowvert \Theta^1 \rangle = -\frac{1}{144 f^4}
\left( -4 (\not K+ \not K') - 11 (\not p + \not p')\right)  \nonumber \\
\langle \Theta^0 \arrowvert t_{mB}^s \arrowvert \Theta^0 \rangle = -\frac{21}{144 f^4}
\left(  (\not K + \not K')-  (\not p +\not p') \right)
\end{eqnarray}
The usual non-relativistic approximation
$\bar{u}\gamma^\mu k_\mu u$ = $k^0$ is applied. Since the $K \pi N$ system is 
bound
by about 30 MeV one can take for a first test $k^0$, $p^0$ as the masses 
of the $K$ and $\pi$
respectively and one sees that the interaction in the I=0 channel is
attractive, while in the I=1 channel is repulsive. 
This would give chances to the $\kappa N$ $t$-matrix to develop a pole in 
the bound region, but rules out the I=1 state. 

The series  of terms of Fig. 4 might lead to a bound state of
$\kappa N$ which would not decay since the only intermediate channel
is made out of $K \pi N$ with mass above the available energy. 
The decay into $K N$ observed experimentally can
be taken into account by explicitly allowing for an intermediate
state provided by   diagrams including $K \pi
\to K \pi$ with the $\pi$ being absorbed by the nucleon in p-wave, which leads
to $K \pi N \to K N$. This and other diagrams accounting for the interaction of
the mesons with the other meson or the nucleon are taken into account in the
calculations \cite{felipe}.

What we find at the end is that, in spite of the attraction found, this
interaction is not enough to bind the system, since we do not find a pole below
the $K 	\pi N$ threshold.
 In order to quantify this second statement 
we increase artificially the potential $t_{mB}$ by adding to it a quantity
which leads to a pole around $\sqrt{s}=1540 \ MeV$ with a width of
around $\Gamma=40 \ MeV$. This is accomplished by 
adding an attractive potential around five or six times bigger than
the existing one.   
This exercise gives a quantitative idea of how far one is from having a
pole. We should however note that we have not exhausted all possible sources of
three body interaction since only those tied to the Weinberg Tomozawa term have
been considered. We think that some  more work in this direction should be 
still encouraged and there are already some steps given in
\cite{Kishimoto:2003xy}.

\section{A resonant $\Delta K$ state as a dynamically generated exotic baryon}
Given the success of the chiral unitary approach in generating dynamically low
energy resonances, one might wander if other resonances could not 
be produced with
different building blocks than those used normally, the octets of stable baryons
and the pseudoscalar mesons.  In this sense, in 
\cite{Kolomeitsev:2003kt} the
interaction of the decuplet of $3/2^+$ with the octet of pseudoscalar mesons is
shown to
lead to many states that have been associated to experimentally well 
established resonances. The purpose of the present work is to show that this
interaction leads also to a new state of positive strangeness, with $I=1$ 
and $J^P =3/2^-$, hence, an exotic baryon which qualifies as a pentaquark in the
quark language, but which is more naturally described in terms of a resonant
state of a $\Delta$ and a $K$.
  
    The lowest order chiral Lagrangian for the interaction of the baryon 
decuplet with the octet of pseudoscalar mesons is given by \cite{Jenkins:1991es}
\begin{equation}
{\cal L}=i\bar T^\mu {\cal D}\!\!\!\!/ T_\mu -m_T\bar T^\mu T_\mu
\label{lag1} 
\end{equation}
where $T^\mu_{abc}$ is the spin decuplet field and $D^{\nu}$ the covariant derivative
given by in \cite{Jenkins:1991es}.

Let us recall the identification of the $SU(3)$ component
of $T$ to the physical states :
$T^{111}=\Delta^{++}$, $T^{112}=\rth\Delta^{+}$, $T^{122}=\rth\Delta^{0}$,
$T^{222}=\Delta^{-}$, $T^{113}=\rth\Sigma^{*+}$, $T^{123}=\rsix\Sigma^{*0}$,
$T^{223}=\rth\Sigma^{*-}$,  $T^{133}=\rth\Xi^{*0}$,
$T^{233}=\rth\Xi^{*-}$, $T^{333}=\Omega^{-}$.

For strangeness $S=1$ and charge $Q=3$ there is only one channel $\Delta^{++}
K^+$ which has $I=2$. For $S=1$ and $Q=2$ there are two channels 
$ \Delta^{++}K^0$ and $\Delta^{+}K^+$. From these one can extract the transition
amplitudes for the $I=2$ and $I=1$
combinations and we find \cite{sarkar}
\begin{equation}
V(S=1,I=2)=\frac{3}{4f^2}(k^0+k^{\prime 0}); ~~~~~V(S=1,I=1)=-\frac{1}{4f^2}
(k^0+k^{\prime 0}),
\label{pot}
\end{equation}
where $k(k^{\prime})$ indicate the incoming (outgoing) meson momenta. 
These results indicate that the
interaction in the $I=2$ channel is repulsive while it is attractive in $I=1$.
This attractive potential and the physical situation is very similar to the one of the
$\bar{K}N$ system in $I=0$, where the interaction is also attractive and leads to
the generation of the $\Lambda(1405)$ resonance 
\cite{Kaiser:1995eg,Oset:1997it,oller,Garcia-Recio:2002td}. 
The use of $V$ as the kernel of the Bethe Salpeter equation
\cite{Oset:1997it}, or the N/D unitary approach of \cite{oller} both lead to the
scattering amplitude 
\begin{equation}
t=(1-VG)^{-1}V
\label{LS}
\end{equation}
 In eq. (\ref{LS}), $V$ factorizes on shell
\cite{Oset:1997it,oller} and $G$ stands for the loop function of the meson and baryon
propagators, the expressions for which are given in \cite{Oset:1997it} for a cut off
regularization and in \cite{oller} for dimensional regularization. 
  
  Next we fix the scale of regularization by determining the cut off, $q_{max}$,
in the loop function of the meson and baryon propagators in order to reproduce 
the resonances for other 
strangeness and isospin channels. They are one resonance in
$(I,S)=(0,-3)$, another one in $(I,S)=(1/2,-2)$ and another one in
$(I,S)=(1,-1)$. The last two appear in \cite{Kolomeitsev:2003kt} around 1800 MeV and 1600 MeV and they are
identified with the $\Xi(1820)$ and $\Sigma(1670)$.  We obtain the same results
as in \cite{Kolomeitsev:2003kt} using a cut off $q_{max}=700$ MeV.

  With this cut off we explore the analytical properties of the amplitude for
$S=1$, $I=1$ in the first and second Riemann sheets. First we see that there is
no pole in the first Riemann sheet. However, if we increase the cut off to 1.5
GeV we find a pole below threshold corresponding to a $\Delta K$ bound state. But
this cut off does not reproduce the position of the resonances discussed above.

  Next we explore the second Riemann sheet for which we take
\begin{equation}
G^{2nd}=G+2i\,\frac{p_{CM}}{\sqrt{s}}\,\frac{M}{4\pi}
\end{equation}
where G is the meson baryon propagator and the variables on the right hand side
 of the equation are evaluated in the first (physical) 
Riemann sheet. In the above equation $p_{CM}$, $M$ and $\sqrt{s}$ denote the CM 
momentum, the $\Delta$ mass and the CM energy respectively.
 We find a pole at $\sqrt{s}=1635$ MeV in the second Riemann
sheet. This should have some repercussion on the physical amplitude and indeed
this is the case as we show below. 
\begin{figure}[tbp]
    \centering
    \includegraphics[width=10cm,clip]{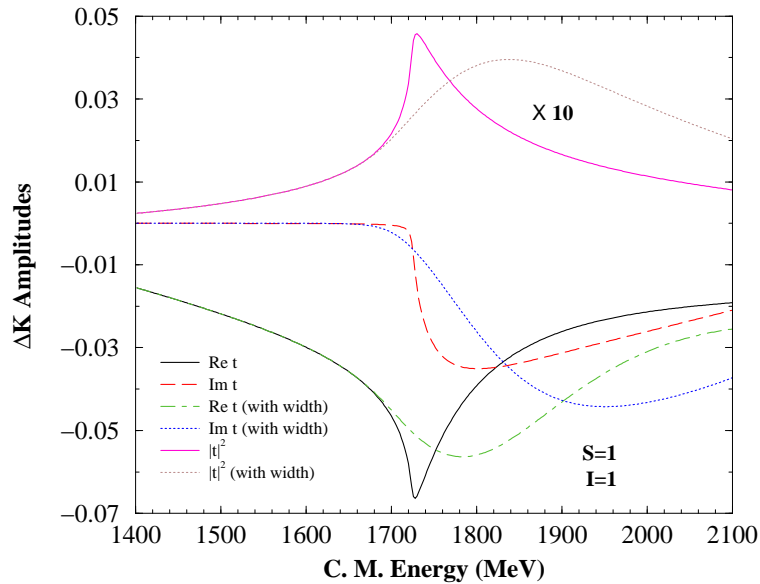}   
\caption{Amplitudes for $\Delta K\rightarrow\Delta K$ for $I=1$}
\end{figure}

\begin{figure}[tbp]
    \centering
    \includegraphics[width=10cm,clip]{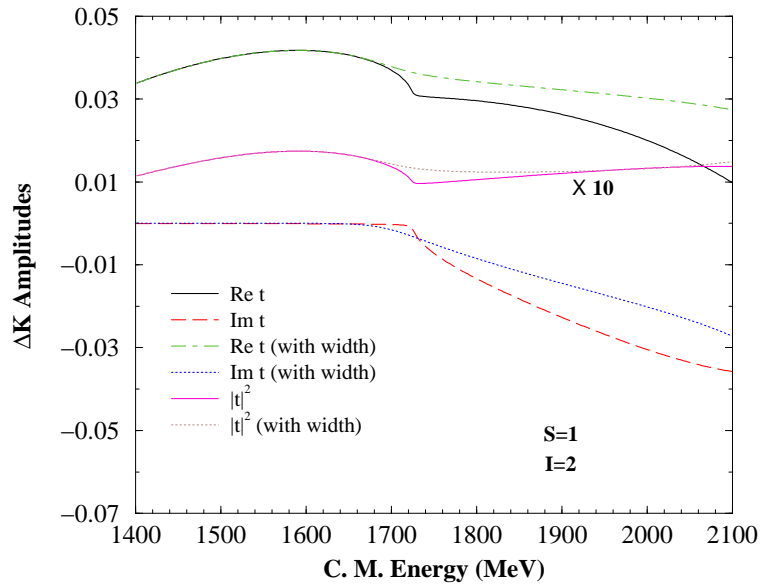}   
\caption{Amplitudes for $\Delta K\rightarrow\Delta K$ for $I=2$}
\end{figure} 

The situation in the scattering matrix is revealed in figs. 5 and 6 which show
the real and imaginary part of the $K \Delta$ 
amplitudes for the case of $I=1$ and $I=2$
respectively. Using the cut off discussed above we can observe  
 the differences between $I=1$ and $I=2$.  For the case of
$I=2$ the imaginary part follows the ordinary behaviour of the opening of a 
threshold,
growing smoothly from threshold. The real part is also smooth, showing
nevertheless the cusp at threshold.
For the case of $I=1$, instead, the strength of the imaginary part is stuck to
threshold as a reminder of the existing pole in the complex plane, growing very
fast with energy close to threshold.  The real part
has also a pronounced cusp at threshold, which is also tied to the same
singularity.  

   We have also done a more realistic calculation taking into account the width
of the $\Delta$ in the intermediate states.   
The results are also shown in figures 5 and 6 and we see that the peaks around
threshold become smoother and some strength is moved to higher energies.  Even
then, the strength of the real and imaginary parts in the $I=1$ are much larger
than for $I=2$. The modulus squared of the amplitudes shows some
peak behavior around 1800 MeV  in the case of $I=1$, while it is small and has no
structure in the case of  $I=2$.

We propose the study 
of the
following reactions:  1) $pp \to \Lambda \Delta^+ K^+$, 2) $pp \to \Sigma^- \Delta^{++}
K^+$, 3) $pp \to \Sigma^0 \Delta^{++}K^0$. In the first case the $\Delta^+ K^+$ state
produced has necessarily $I=1$.  In the second case the $\Delta^{++}K^+$ state has
$I=2$. In the third case the $\Delta^{++}K^0$ state has mostly an $I=1$ component.
The study of these reactions, particularly the invariant mass distribution of 
$\Delta K$, and the comparison of the $I=1$ and
$I=2$ cases would provide the information we are searching for.  Indeed, the mass
distribution is given by
\begin{equation}
\frac{d \sigma}{dm_{I(\Delta K)}} = C |t_{\Delta K \rightarrow \Delta K}|^2
p_{CM}
\label{sig}
\end{equation}
where $p_{CM}$ is the $K$ momentum in the $\Delta K$ rest 
frame. The mass distribution removing the  $p_{CM}$ factor in eq. (\ref{sig}) should 
show the broad peak of $|t_{\Delta K \rightarrow \Delta K}|^2$ seen in fig. 5.  Similarly, the ratio of mass
distributions in the cases 3) to 2) or 1) to 2), discussed before, should show 
this behaviour.

Given the success of the chiral unitary approach providing dynamically generated
resonances in the interaction of the octet of $1/2^+$ baryons with the octet of
pseudoscalar mesons, as well as in the scalar sector of the meson meson
interaction \cite{report}, the predictions made here stand on firm ground. The
experimental confirmation of the results found here would give evidence for
another pentaquark state which, however, stands for a simple description as a
resonant $\Delta K$ state. 

\section{Acknowledgments}
This work is partly supported by DGICYT contract number BFM2003-00856,
 the E.U. EURIDICE network contract no. HPRN-CT-2002-00311 and the Research
 Cooperation program of the japanese JSPS and the spanish CSIC.

\end{document}